\documentclass{jps-cp}
\usepackage{txfonts} %Please comment out this line unless the txfonts package is availabe in your LaTeX system.

\title{Theory of Muon g-2}

\author{Bogdan \textsc{Malaescu}$^{1}$}

\inst{$^{1}$LPNHE, Sorbonne Universit\'e, Universit\'e Paris Cit\'e, CNRS/IN2P3, Paris, France}

\email{malaescu@in2p3.fr}

\recdate{March 15, 2022}

\abst{The longstanding discrepancy between the measured and the predicted values of the anomalous magnetic moment of the muon, $a_\mu = (g-2)/2$, is one of the most intriguing potential hints of new physics in particle physics. After a brief introduction, the status of the theoretical prediction of g-2 is presented, with some focus on the contributions yielding the dominant uncertainties. The status of the comparison with the experimental measurement is then discussed.}

\kword{muon, g-2, prediction}

\begin{document}
\maketitle

\newcommand{\epem}{\mathrm{e^+e^-}}
\def\gev{\mathrm{Ge\kern-0.1em V}}

\newcommand{\amuSM}{\ensuremath{a_\mu^\text{SM}}}

\newcommand{\amuexp}{\ensuremath{a_\mu^\text{exp}}}

\newcommand{\amuHVP}{\ensuremath{a_\mu^\text{HVP}}}

\newcommand{\amuHVPLO}{\ensuremath{a_\mu^\text{HVP, LO}}}
\newcommand{\amuHVPLOpp}{\ensuremath{a_\mu^\text{HVP, LO}[\pi\pi]}}
\newcommand{\amuHVPLOppp}{\ensuremath{a_\mu^\text{HVP, LO}[3\pi]}}

\newcommand{\amuHVPNLO}{\ensuremath{a_\mu^\text{HVP, NLO}}}
\newcommand{\amuHVPNNLO}{\ensuremath{a_\mu^\text{HVP, NNLO}}}

\newcommand{\amuHLbL}{\ensuremath{a_\mu^\text{HLbL}}}
\newcommand{\amuHLbLNLO}{\ensuremath{a_\mu^\text{HLbL, NLO}}}

\newcommand{\amuhad}{\ensuremath{a_\mu^\text{had}}}
\newcommand{\amuQED}{\ensuremath{a_\mu^\text{QED}}}

\newcommand{\amuW}{\ensuremath{a_\mu^\text{Window}}}

\section{Introduction}

A longstanding discrepancy has been observed between the measured and the predicted values of the anomalous magnetic moment of the muon, $a_\mu = (g-2)/2$.
This represents a potential hint of new physics in particle physics and calls for an improvement of the theoretical prediction, in view of the significant gain in precision foreseen for the Fermilab measurement.
A series of workshops organised by the Muon g-2 Theory Initiative has allowed for detailed discussions on the various approaches that are being considered for computing the different contributions to $a_\mu$.
A consensus has been reached for a conservative evaluation of the theoretical prediction and its uncertainty.
The results presented here are based mainly on the White Paper of the Muon g-2 Theory Initiative~\cite{Aoyama:2020ynm} and on the inputs used therein from Refs.~\cite{Davier:2017zfy,Keshavarzi:2018mgv,Colangelo:2018mtw,Hoferichter:2019mqg,Davier:2019can,Keshavarzi:2019abf,Kurz:2014wya,FermilabLattice:2017wgj,Budapest-Marseille-Wuppertal:2017okr,RBC:2018dos,Giusti:2019xct,Shintani:2019wai,FermilabLattice:2019ugu,Gerardin:2019rua,Aubin:2019usy,Giusti:2019hkz,Melnikov:2003xd,Masjuan:2017tvw,Colangelo:2017fiz,Hoferichter:2018kwz,Gerardin:2019vio,Bijnens:2019ghy,Colangelo:2019uex,Pauk:2014rta,Danilkin:2016hnh,Jegerlehner:2017gek,Knecht:2018sci,Eichmann:2019bqf,Roig:2019reh,Colangelo:2014qya,Blum:2019ugy,Aoyama:2012wk,Aoyama:2019ryr,Czarnecki:2002nt,Gnendiger:2013pva}.

While $a_\mu$ is sensitive to quantum fluctuations of fields, it has to be computed very precisely, which is a highly-nontrivial exercise.
This evaluation involves QED contributions, currently known up to O($\alpha^5$)~\cite{Aoyama:2012wk,Aoyama:2019ryr}, which corresponds to a precision of $0.001$~ppm.
Then there are electroweak contributions, currently known with a precision of $0.01$~ppm.
The dominant uncertainties originate from non-perturbative contributions, namely the hadronic vacuum polarisation~(HVP) and the hadronic light-by-light~(HLbL), with a precision of $0.34$~ppm and $0.15$~ppm respectively.
All these contributions have to be well under control before one can make any claim about the observation of potential contributions from new physics, beyond the Standard Model.

\section{Status of the theoretical prediction}

The dominant uncertainty for the theoretical prediction originates from lowest-order HVP contribution~(\amuHVPLO), which cannot be calculated based on perturbative QCD because of the low mass scale.
Instead, one can use experimental data on hadronic production cross section~( $\sigma \left( \epem \rightarrow {\rm hadrons} \right) $ ).
Indeed, the optical theorem relates the imaginary part of the two-point correlator to the hadronic cross section and a dispersion integral allows to compute \amuHVPLO.
The kernel of this integral strongly enhances the low energy region and therefore precise measurements of the $\sigma \left( \epem \rightarrow {\rm hadrons} \right) $ in this region are very important.
We do not use anymore data from hadronic $\tau$ decays, because they are less precise and are also impacted by theory uncertainties~\cite{Davier:2019can}.

\begin{figure}[tb]
\includegraphics[width=7.5cm]{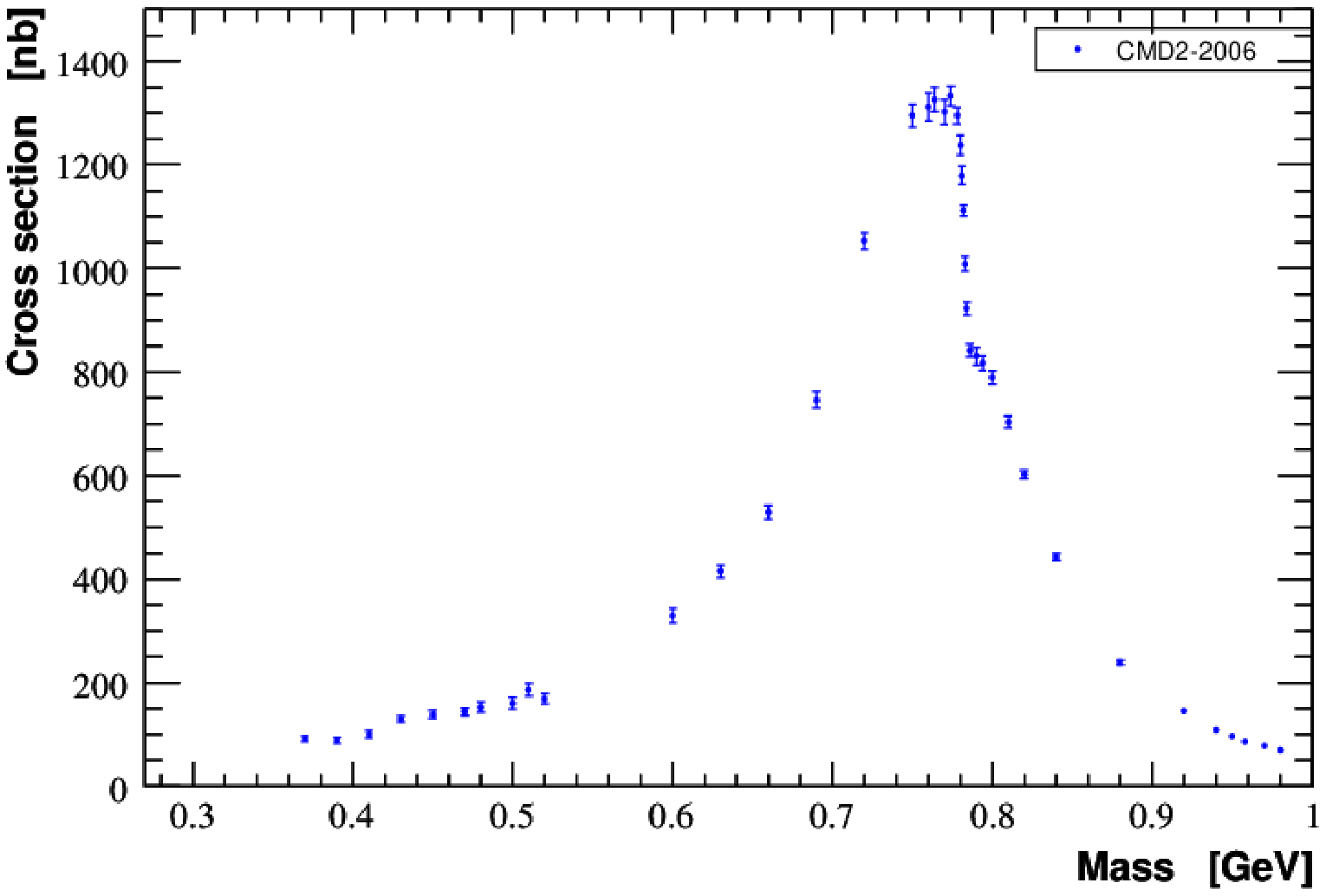}
\hspace{0.1cm}
\includegraphics[width=7.5cm]{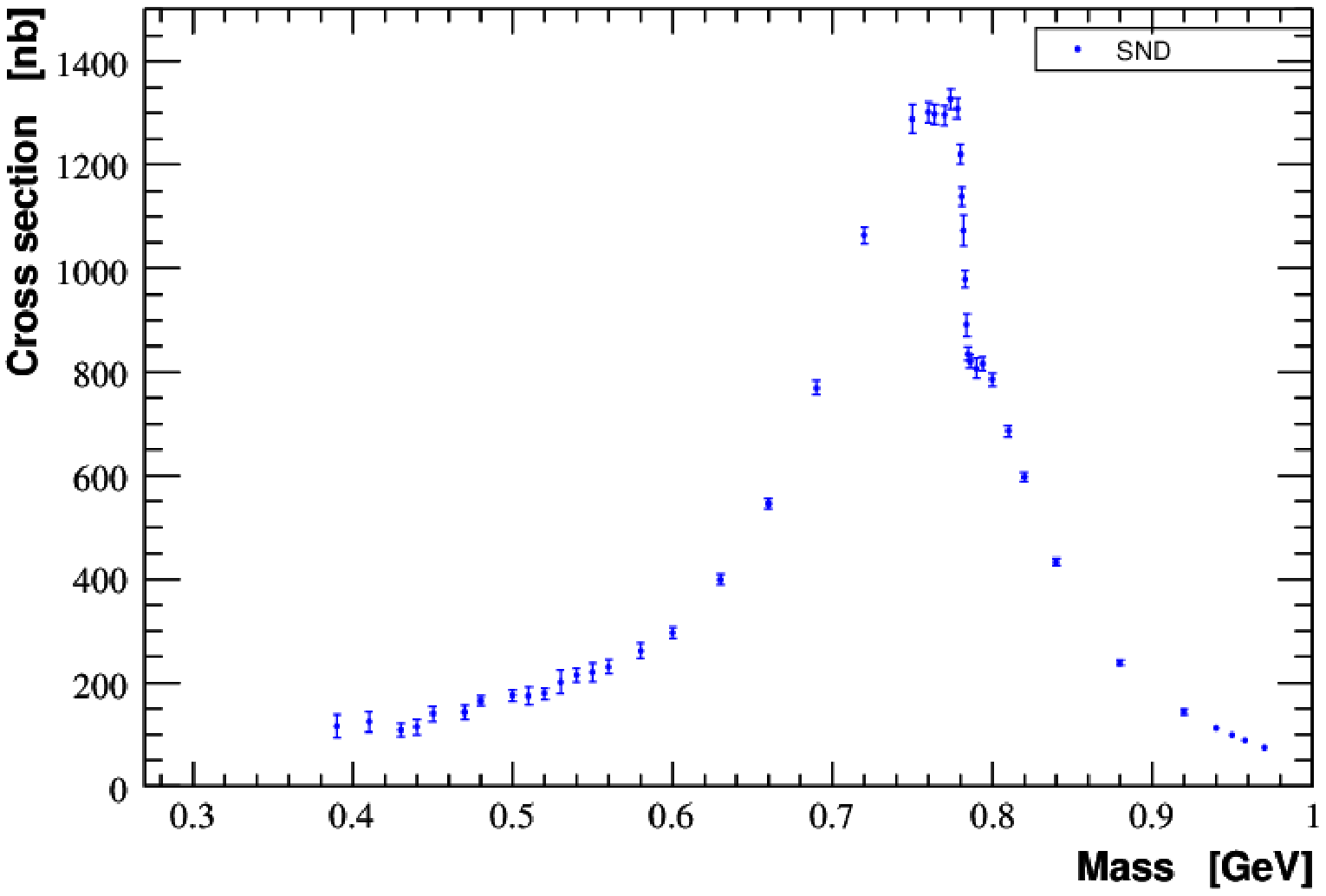} \\
\includegraphics[width=7.5cm]{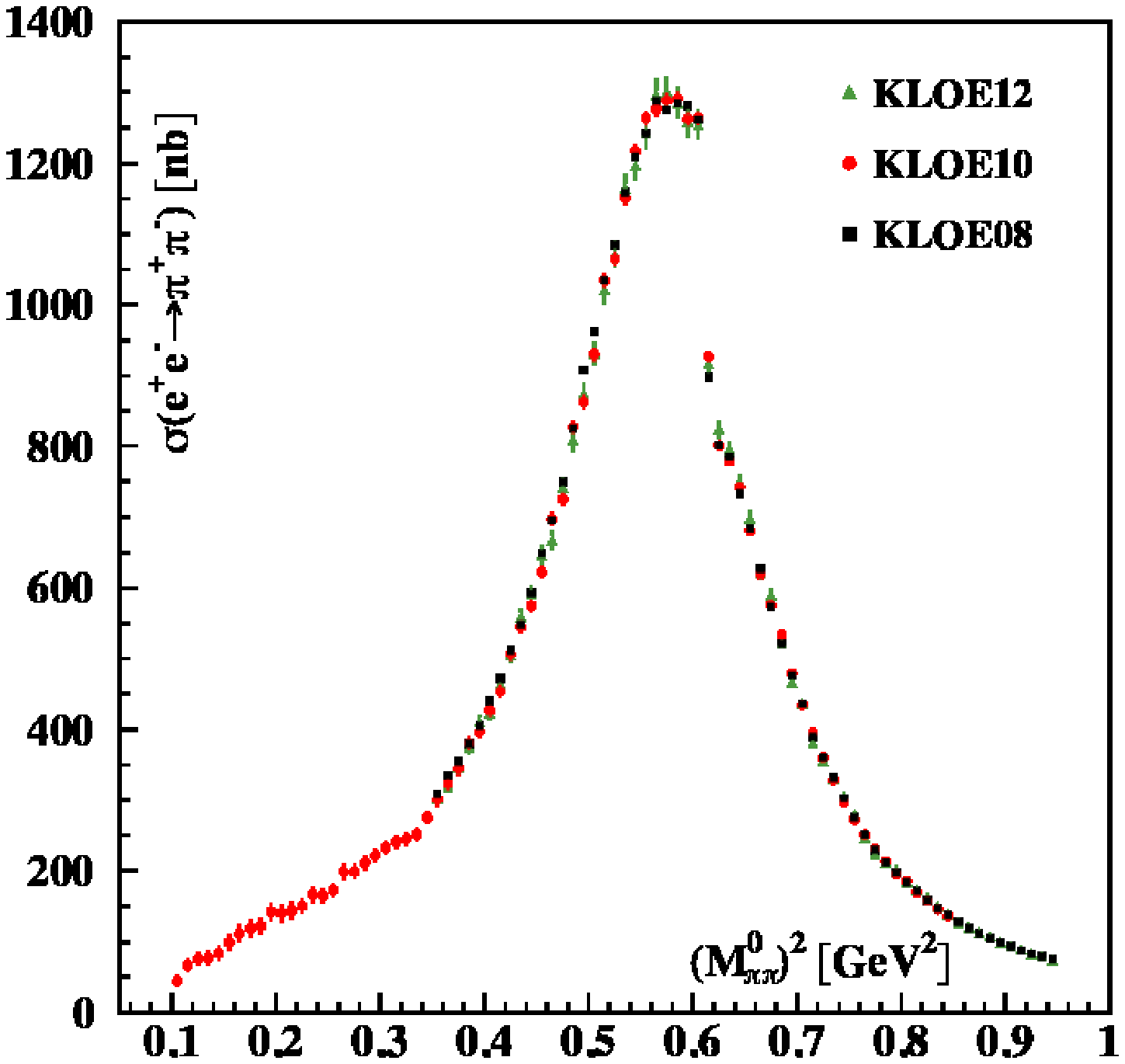}
\hspace{0.1cm}
\includegraphics[width=7.5cm]{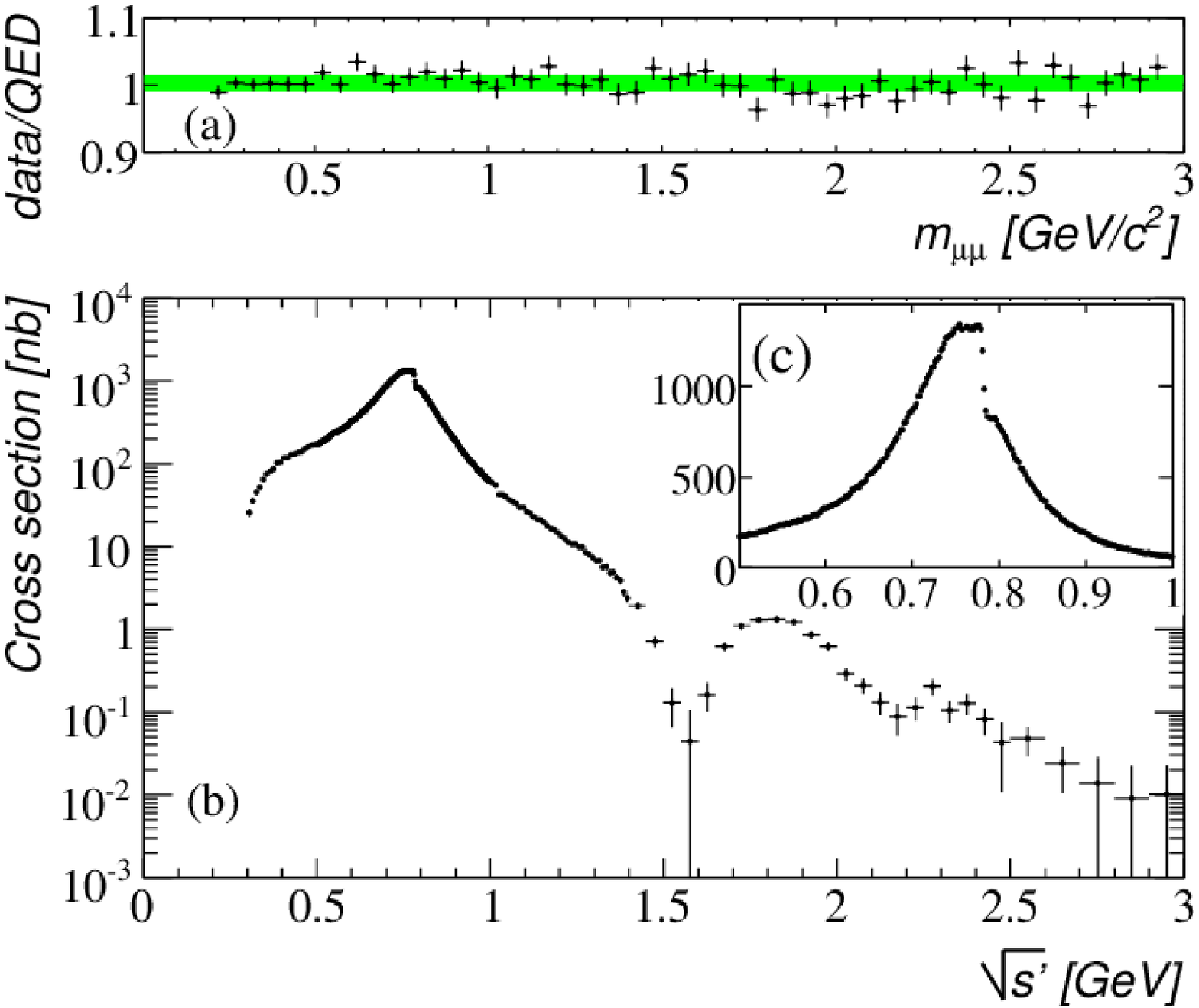} 
\caption{The top plots show data from CMD-2~\cite{Aulchenko:2006dxz,CMD-2:2006gxt} (left) and SND~\cite{Achasov:2006vp} (right) on $\epem \rightarrow \pi^+\pi^-$ in the $\rho$ region.
Bottom-left: the plot, adapted from Ref.~\cite{KLOE:2012anl}, shows the KLOE data sets on $\epem \rightarrow \pi^+\pi^-$ in the $\rho$ region obtained in the three experimental configurations described in the text~\cite{KLOE:2008fmq,KLOE:2010qei,KLOE:2012anl}.
Bottom-right: the plot, reprinted from Ref.~\cite{BaBar:2009wpw}, displays the results from BABAR~\cite{BaBar:2009wpw,BaBar:2012bdw} using the large-angle ISR method: $\epem \rightarrow \mu^+\mu^-$ compared to NLO QED (top frame) and  $\epem \rightarrow \pi^+\pi^-$ from threshold to $3\gev$ using the $\pi\pi/\mu\mu$ ratio (bottom frame). The insert shows the $\rho$ region.}
\label{Fig:ExpData}
\end{figure}

Fig.~\ref{Fig:ExpData} shows examples of experimental spectra for the dominant $\pi^+\pi^-$ channel.
At CMD-2~\cite{Aulchenko:2006dxz,CMD-2:2006gxt} and SND~\cite{Achasov:2006vp} the measurements are performed through a scan of the energy in the centre-of-mass of the collider.
KLOE used the ISR technique, selecting events with a hard photon emitted from the initial state, which allows to cover a broad range of masses for the hadronic final state system~\cite{KLOE:2008fmq,KLOE:2010qei,KLOE:2012anl}. 
The measurement performed at BABAR using the ISR approach selects events with the hard photon reconstructed in the detector, allowing to cover the full energy range of interest.
Furthermore, the ISR luminosity is evaluated in-situ based on $\epem \rightarrow \mu^+\mu^-$ events, which allows to reduce several systematic uncertainties, and in both the $\pi\pi$ and $\mu\mu$ channels events with extra photons are considered~(performing an "NLO measurement").
In the $\mu\mu$ channel, the comparison of the measured spectrum with the NLO QED-based prediction provides an excellent test of the procedure~\cite{BaBar:2009wpw,BaBar:2012bdw}.
While the KLOE $'08$~\cite{KLOE:2008fmq} and KLOE $'12$~\cite{KLOE:2012anl} measurements consider events with the hard ISR photon along the beam, the KLOE $'10$ measurement uses large-angle ISR photons~\cite{KLOE:2010qei}.
The KLOE $'12$ measurement also uses the muon spectrum for the normalisation~\cite{KLOE:2012anl}.
Still, for the time being, only the BABAR study uses the reconstructed extra photons, necessary at the sub-percent level precision aimed for here.

\begin{figure}[tb]
\includegraphics[width=7.5cm]{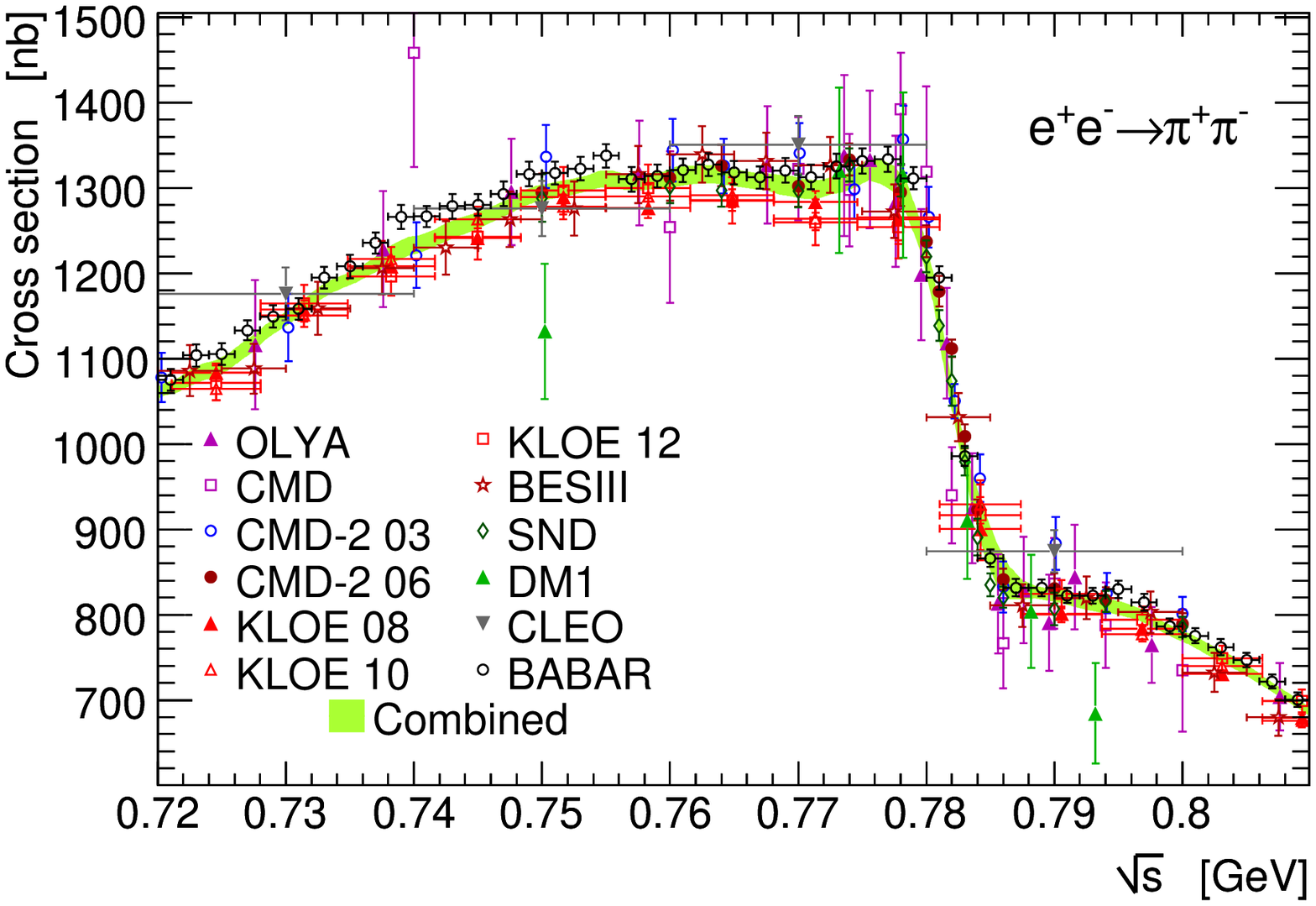}
\hspace{0.1cm}
\includegraphics[width=7.5cm]{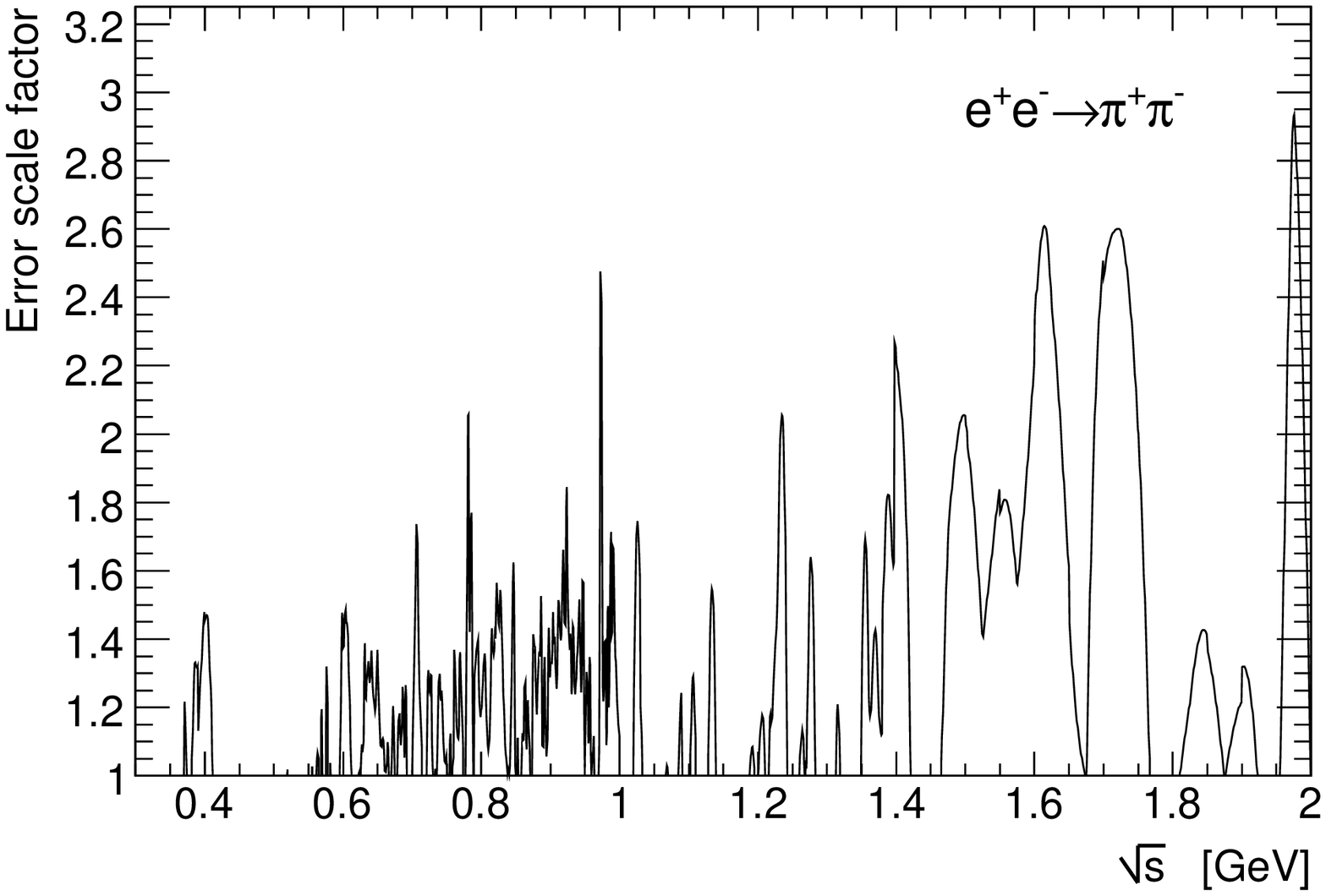}
\caption{Left: Cross section for $\epem\to \pi^+\pi^-$ annihilation measured by the different experiments for the entire energy range. The error bars contain both statistical and systematic uncertainties, added in quadrature. The shaded (green) band represents the average of all the measurements obtained by HVPTools, which is used for the numerical integration following the procedure discussed in Ref.~\cite{Davier:2019can}. Right: Rescaling factor accounting for inconsistencies among experiments versus $\sqrt{s}$. Reprinted from Ref.~\cite{Davier:2019can}.}
\label{Fig:comb}
\end{figure}

Computing the \amuHVPLO~contribution requires combining the experimental spectra measured by various experiments, as performed e.g. by the DHMZ~\cite{Davier:2017zfy,Davier:2019can}~(see Fig.~\ref{Fig:comb} left) and KNT~\cite{Keshavarzi:2018mgv,Keshavarzi:2019abf} teams respectively, and summing the contributions of the various exclusive channels.
While the channel-by-channel combinations are performed taking into account the information on the uncertainties and their correlations between bins/points and between experiments, the DHMZ approach~(implemented in the HVPTools software) also accounts for correlations between different channels.
This induces a necessary enhancement of the total uncertainty.

Furthermore, a method for taking into account the local tensions between the measurements~(based on the computation of a $\chi^2/{\rm ndof}$ in fine energy ranges, used for the rescaling of the uncertainties) has been implemented by DHMZ in Refs.~\cite{Davier:2010rnx,Davier:2010nc} and is still being used~\cite{Davier:2017zfy,Davier:2019can}~(see Fig.~\ref{Fig:comb} right).
Such approach has also been used in the more recent KNT studies~\cite{Keshavarzi:2018mgv,Keshavarzi:2019abf}.
Still, this local rescaling of the uncertainties is not sufficient in presence of systematic tensions, as the ones between the BABAR and the KLOE measurements in the $\pi\pi$ channel.
These tensions are taken into account in the most recent DHMZ study~\cite{Davier:2019can} by treating half of the difference between integrals without either BABAR or KLOE as an extra uncertainty.
This yields the dominant uncertainty in the study, amounting to $2.8 \cdot 10^{-10}$.

\begin{figure}[tb]
\includegraphics[width=7.5cm]{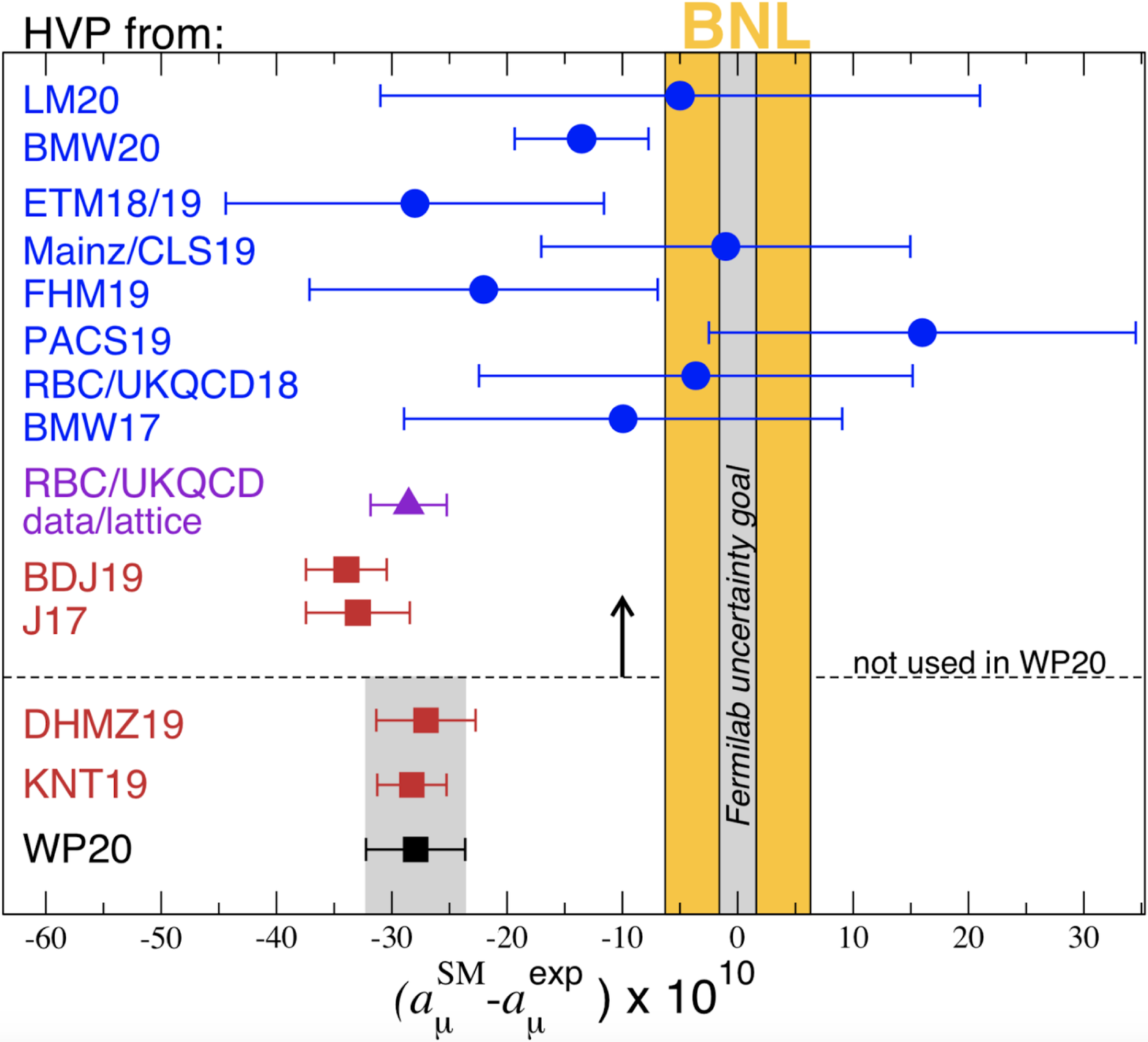}
\hspace{0.1cm}
\includegraphics[width=7.5cm]{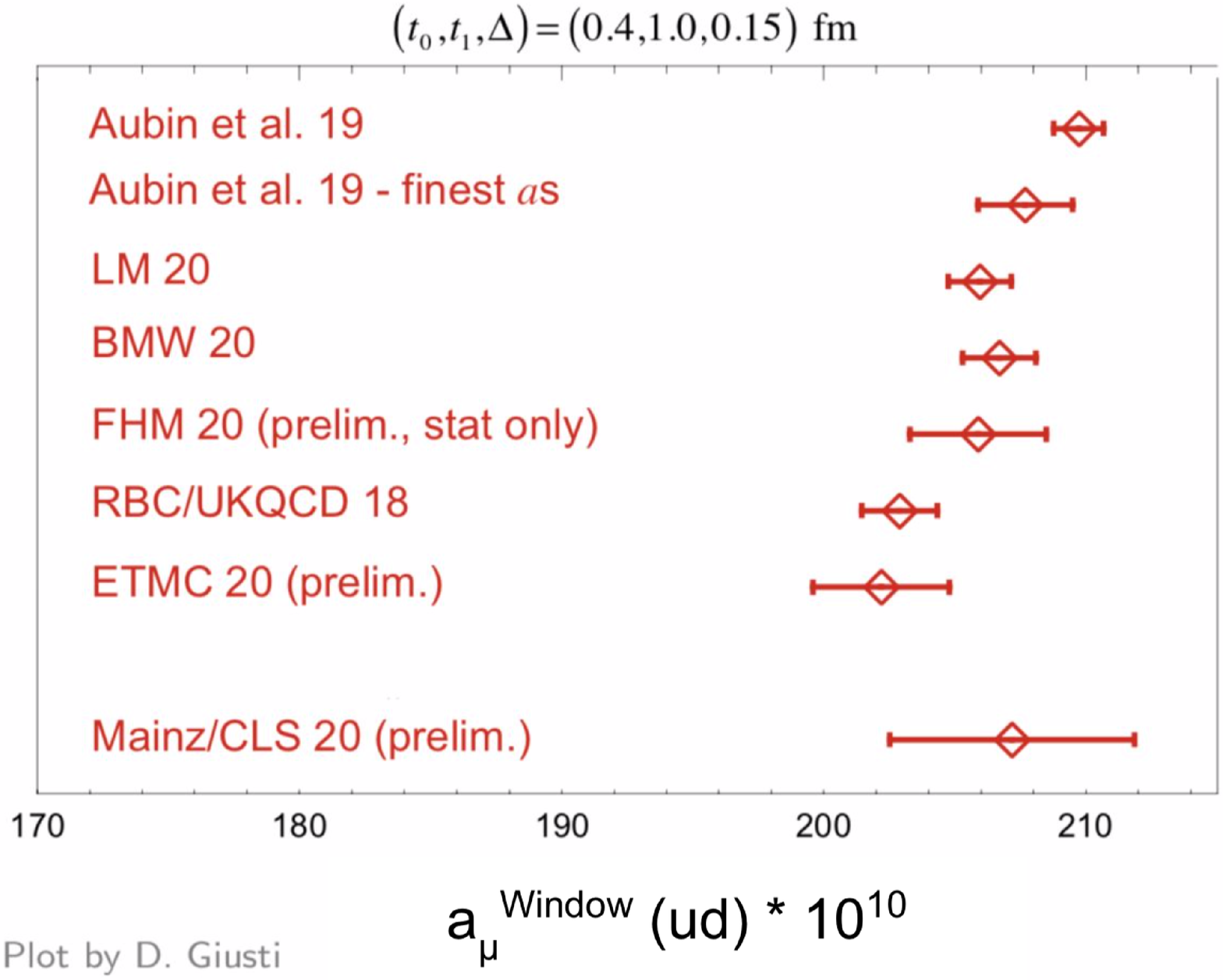}
\caption{Left: The theoretical predictions based on various determinations of \amuHVPLO~are compared with the BNL experimental result~\cite{Muong-2:2006rrc}, indicated by the yellow vertical band. The result of the Theory Initiative White Paper~\cite{Aoyama:2020ynm}~(WP20), as well as the DHMZ19~\cite{Davier:2019can} and KNT19~\cite{Keshavarzi:2019abf} inputs, are indicated below the dashed horizontal line. The uncertainty goal for the Fermilab measurement is also displayed. 
Right: Comparison of the values for a "Window" moment defined in Ref.~\cite{RBC:2018dos}, for various Lattice QCD studies.}
\label{Fig:comparison}
\end{figure}

Fig.~\ref{Fig:comparison} left shows the comparison between the theoretical predictions based on various determinations of \amuHVPLO~and the BNL experimental result~\cite{Muong-2:2006rrc}.
The result of the Theory Initiative White Paper~\cite{Aoyama:2020ynm} is obtained through the merging of model independent results, as obtained by DHMZ~\cite{Davier:2019can} and KNT~\cite{Keshavarzi:2019abf}~(and CHHKS for the $\pi^+\pi^-$~\cite{Colangelo:2018mtw} and $\pi^+\pi^-\pi^0$~\cite{Hoferichter:2019mqg} channels). 
The corresponding central value is obtained from the simple average of the input integrals.
The largest of the DHMZ and KNT uncertainties is used in each channel.
The treatment of the BABAR-KLOE tension in the $\pi\pi$ channel and of the correlations between the various channels is based on the DHMZ result.
The importance of properly accounting for these two aspects, as done in the DHMZ study, is indeed visible when comparing the uncertainties of the DHMZ and KNT results~(see Fig.~\ref{Fig:comparison} left).

During the last few years there has been excellent progress on the Lattice QCD (+QED) calculations~\cite{FermilabLattice:2017wgj,Budapest-Marseille-Wuppertal:2017okr,RBC:2018dos,Giusti:2019xct,Shintani:2019wai,FermilabLattice:2019ugu,Gerardin:2019rua,Aubin:2019usy,Giusti:2019hkz,Borsanyi:2020mff}.
In particular, while the BMW20~\cite{Borsanyi:2020mff} result is still to be cross-checked by other lattice groups, its precision became similar to the one of dispersive approaches.
There are indeed ongoing cross-checks using Euclidean time windows~(related to the HVP with suppression of very low and high energies~\cite{RBC:2018dos}) for which various groups achieved similar precision~(see Fig.~\ref{Fig:comparison} right).
If the BMW20 result is confirmed, the difference with respect to the dispersive results will have to be understood.

\begin{figure}[tb]
\begin{center}
%\noindent\begin{minipage}{1.5\textwidth}
\includegraphics[width=7.5cm]{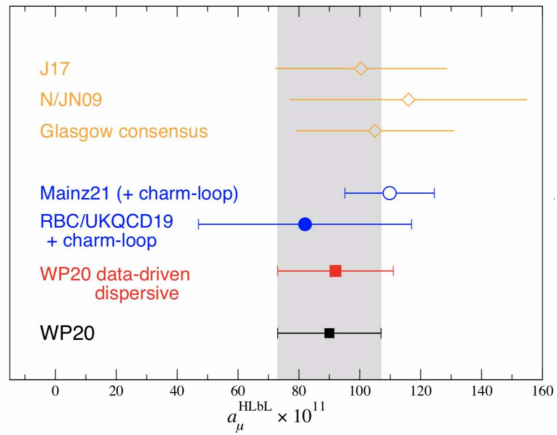}
%\end{minipage}
\end{center}
\caption{Summary of the \amuHLbL~values obtained through various approaches: hadronic model + perturbative QCD~(yellow diamond), lattice QCD+QED~(blue circles) and data-driven~(red square). The result of the Theory Initiative White Paper~\cite{Aoyama:2020ynm}~(WP20) indicated by the vertical grey band uses only the inputs marked with filled symbols.}
\label{Fig:HLbL}
\end{figure}

A summary of the \amuHLbL~values obtained through various approaches, as well as the result of the Theory Initiative White Paper~\cite{Aoyama:2020ynm}, are indicated in Fig.~\ref{Fig:HLbL}.
One can notice that there is indeed good progress on systematically improvable approaches.
Furthermore, the g-2 Theory Initiative provides an adequate environment for cross-checks among these results.
The uncertainty of \amuHLbL~is currently controlled with a precision of $0.15$~ppm.
The result obtained by the Mainz group~\cite{Chao:2020kwq}, with a somewhat better precision than the result of the Theory Initiative White Paper~\cite{Aoyama:2020ynm}, became publicly available after the latter and is to be considered in future iterations.

\begin{figure}[tb]
\begin{center}
%\noindent\begin{minipage}{1.5\textwidth}
\includegraphics[width=9.cm]{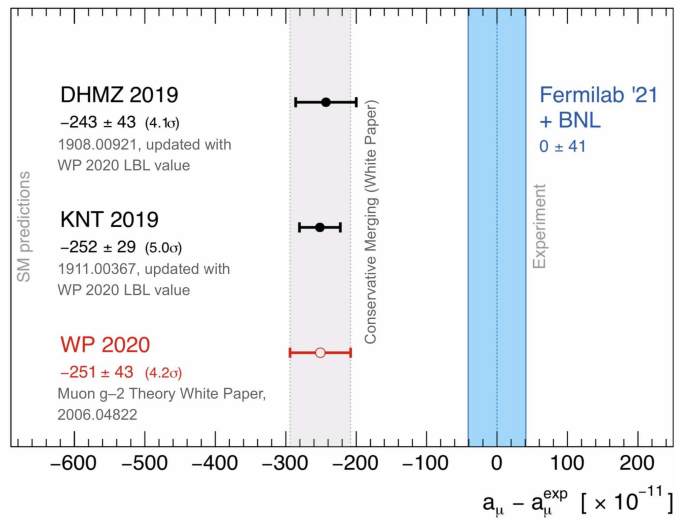}
%\end{minipage}
\end{center}
\caption{The theoretical predictions based on various data-driven determinations of \amuHVPLO~are compared with the Fermilab~\cite{Muong-2:2021ojo} and BNL~\cite{Muong-2:2006rrc} combined experimental result, indicated by the blue vertical band. The result of the Theory Initiative White Paper~\cite{Aoyama:2020ynm} is indicated by the red empty circle and the gray vertical band. The DHMZ19~\cite{Davier:2019can} and KNT19~\cite{Keshavarzi:2019abf} results, updated to use the \amuHLbL~from the White Paper, are indicated by the filed black circles.}
\label{Fig:DispHVPcomp}
\end{figure}

Putting together all the contributions to the theoretical prediction, the dominant uncertainty originates from \amuHVPLO, based on the merging of model-independent methods, while the \amuHLbL~is impacted by an important uncertainty too~\cite{Aoyama:2020ynm}.
At the same time, Lattice QCD results become more and more interesting.
In Fig.~\ref{Fig:DispHVPcomp} the theoretical predictions based on various data-driven determinations of \amuHVPLO are compared with the Fermilab~\cite{Muong-2:2021ojo} and BNL~\cite{Muong-2:2006rrc} combined experimental result.
A tension between the BNL measurement and the reference SM prediction, at the level of $3.7~\sigma$ is observed.
The tension increases to $4.2~\sigma$ when including the FNAL result.
The tension is significantly smaller when using BMW20 for the \amuHVPLO, a result which is still to be confirmed by other lattice groups.

% Label figures, tables, and equations appropriately using the \verb|\label| command, and use the \verb|\ref| command to cite them in the text as ``\verb|as shown in Fig. \ref{Fig:comb}|". This automatically labels the numbers in numerical order.
% The \verb|minipage| environment can be used to place figures horizontally.

% \begin{equation}
% E = mc^{2}
% \label{e1}
% \end{equation}

% \appendix
% \section{}
% Use the \verb|\appendix| command if you need an appendix(es). The \verb|\section| command should follow even though there is no title for the appendix (see above in the source of this file).

\section{Conclusion}

While the tension between the nominal Standard Model prediction and the experimental measurement of $a_\mu$ has reached the level of $4.2$ standard deviations, it is worth phrasing some words of caution about this significance.
Indeed, while the experimental measurement is statistics-dominated, the uncertainty of the prediction is limited by non-Gaussian systematic effects that do not have a rigorous statistical interpretation~(see also the discussion in Ref.~\cite{Cowan:2021sdy}).
Nevertheless, one can clearly state that there is a large discrepancy between measurement and reference Standard Model prediction.
The latter needs to be significantly improved in view of the forthcoming updates of the Fermilab measurement.
One can therefore conclude that we have an interesting, long standing, multifaceted problem to solve.

\section*{Acknowledgements}

I kindly thank the organisers of the SPIN 2021 Symposium for the invitation to give this talk.
I thank my colleagues and friends Michel DAVIER, Andreas HOECKER and Zhiqing ZHANG for our fruitful collaboration, as well as the members of the g-2 Theory Initiative for the numerous interesting discussions.

\end{document}